
\magnification = 1200
\tolerance 1000
\pretolerance 1000
\baselineskip 20pt
\overfullrule = 0pt
\hsize=16.3truecm
\def\sp{\kern 1em}

\centerline {\bf  RESONANT ANDREEV SCATTERING}
\centerline {\bf IN PHASE-COHERENT, SUPERCONDUCTING
NANOSTRUCTURES.  }
\vskip 0.5truecm

\centerline {\bf N.R. Claughton, M. Leadbeater, C.J. Lambert and
V.N. Prigodin$^\dagger$}

\centerline {\bf School of Physics and Chemistry}

\centerline {\bf Lancaster University}

\centerline {\bf Lancaster   LA1 4YB}

\centerline {\bf U.K.}

$^\dagger${\bf Max Planck Institute, Stuttgart, Germany}

\noindent
{\bf ABSTRACT.}

Analytic predictions for resonant transport in three generic structures are
presented. For a structure comprising a normal (N) contact - normal dot (NDOT)
-
superconducting (S) contact, we predict that finite voltage, differential
conductance resonances are destroyed by the switching on of superconductivity
in
the S-contact. In the weak coupling limit, the  surviving resonances  have a
double-peaked line-shape. Secondly, we demonstrate that resonant Andreev
interferometers can provide galvonometric magnetic flux detectors, with a
sensitivity in excess of the flux quantum. Finally, for a superconducting dot
(SDOT) connected to normal contacts (N), we show that the onset of
superconductivity  can increase the sub-gap conductance, in contrast with the
usual behaviour of a tunnel junction.


\vfil\eject
Recent advances in the fabrication of nanoscale structures
have led to increasing interest in
 transport through resonant tunnel junctions and quantum dots[1,2].
 In part this
is due to the new physics associated with Coulomb blockade[3] and in part due
to growing interest in quantum chaos[4-7]. The bulk of work in this area has
focussed on normal structures, but more recently attention has turned
to hybrid structures involving a superconducting component. It has been
demonstrated experimentally that the energy gap of a superconducting dot
is directly observable through Coulomb blockade experiments[8] and
theoretical work on incoherent transport through such dots has been
carried out[9].
In this Letter, we describe the effect of
superconductivity on phase-coherent transport
through resonant structures, in the limit that charging effects
can be ignored. This limit should be experimentally accessible,
because even intimate contact with a superconductor will not
broaden states below the gap.

A range of new phenomena and fundamental
problems involving coherent
transport through resonant superconducting hybrids
are expected to manifest themselves in a small number of
generic structures.
One such example is a \lq\lq N-NDOT-S" structure
comprising one or more normal (N) current carrying leads,
 in contact with a normal zero dimensional \lq\lq dot" (NDOT), which in turn
 makes contact with a superconducting (S) lead.
In contrast with the zero voltage limit, where
a general multi-channel description of this structure is available[10],
there currently exists only
a single one-dimensional study of finite voltage,
resonant transport[11], in which
 a $\delta$-function potential well, with a
single localized state is introduced into a one-dimensional
 insulating barrier.
A key prediction of this Letter is that finite-voltage conductance resonances
are destroyed by the onset of superconductivity in such a structure.
This is illustrated in figures 1 and 2, which show for a
two-dimensional system, the conductance
of a resonant structure plotted against the Fermi energy of the dot.
Figures 1a and 2a show the conductance of
a N-NDOT-N structure at zero and finite voltages respectively, while
figures 1b and 2b show the conductance of the corresponding N-NDOT-S
structure.

A second class of  structures involves two (or more)
separate superconductors S and S$^\prime$,
with respective order parameter phases $\phi$, $\phi'$.
The resonant energies of a N-SS$^\prime$ or
N-NDOT-SS$^\prime$
composite will vary periodically with the
phase difference $\phi-\phi'$[15-26] and in what follows we
describe the resonant properties of such structures.
A third class is formed
when a normal lead makes contact with a superconducting dot (SDOT),
which in turn makes contact with a second normal lead.
In such structures, we demonstrate that the
sub-gap conductance can increase when superconductivity is switched on,
in contrast with the conventional behaviour of a N-S tunnel junction.

In what follows, we adopt a multi-channel scattering
approach, based on
current-voltage relations for  phase-coherent scatterers
written down in reference [27], which
have been extended and re-derived in several papers [28-30].
In the absence of inelastic scattering,
dc transport is determined by the
quantum mechanical scattering matrix $s(E,H)$, which yields scattering
properties of quasi-particles of energy $E$, incident on
 a phase-coherent structure
described by a Hamiltonian $H$.
If the structure is connected to external reservoirs by open
scattering channels labelled by quantum numbers $n$, then this has matrix
elements of the form $s_{n,n'}(E,H)$. The squared modulus of  $s_{n,n'}(E,H)$
is the outgoing
flux of quasi-particles along channel $n$, arising from a unit incident
flux along channel $n'$.
For channels belonging to current-carrying leads, with quasi-particles
labelled by a discrete quantum number $\alpha$
($\alpha = + 1$ for particles, $-1$ for holes),
it is convenient to
write $n=(i,a,\alpha)$, where $i$ labels the leads
and $a$ labels all other quantum numbers
associated with the channels.
Then as shown in [27,28], transport properties are determined by
the quantity
$$ P^{\alpha,\beta}_{i,j}(E,H)=\sum_{a,b}\vert
s_{(i,a),(j,b)}^{\alpha,\beta}(E,H)
\vert^2, $$
which is referred to as either a reflection probability ($i=j$) or a
transmission
probability ($i\ne j$) from quasi-particles of type $\beta$ in lead $j$ to
quasi-particles of type $\alpha$ in lead $i$. For $\alpha \ne\beta$,
$ P^{\alpha,\beta}_{i,j}(E,H)$ is referred to as
an Andreev scattering probability, while for
$\alpha=\beta$, it is a normal scattering probability.

The starting point for our description of resonant Andreev scattering is the
following general formula for the transition probability between two
different scattering channels of an open vector space $A$ representing the
leads,
attached to a closed sub-space $B$ representing the scatterer.
For $n\ne n'$, the result,
which we derive[31] for normal leads described by a real
Hamiltonian, is
$$T_{nn'}=4\,{\rm Trace}\left[{\Gamma}(n){ G_{BB}}{\Gamma}(n')
{ G_{BB}}^\dagger\right]\eqno{(1)},$$
where
$${ G_{BB}}^{-1}={ g_B}^{-1}-{\sigma'}-{\sigma}+\imath{\Gamma}
\eqno{(2)}$$
and the trace is over all internal levels of  $B$.
In these expressions,
${ \Gamma}(n)$ is a Hermitian matrix of inverse lifetimes,
${ \Gamma}=\sum_n{ \Gamma}(n)$,
${ \sigma}$ and ${ \sigma'}$ are Hermitian self-energy matrices and
${ g_{B}}$ is the retarded Green's function of sub-space $B$ when ${ H_1}=0$.
The above result has been cast in a form which resembles
the Breit-Wigner formula [12,13], but is
 very general and makes no assumptions about the presence
or otherwise of resonances.

 During the past decade, the Breit-Wigner formula
has been applied to a variety of problems involving resonant transport in
normal-state structures [32-34].  For a normal-metallic conductor,
under resonant conditions, where the level spacing is much greater
than the broadening, B\"uttiker has presented
a multi-channel derivation of the Breit-Wigner formula through a single
resonant level [35].
This limit is recovered from equation (1)
by  restricting the trace to a single level.
In what follows, we shall encounter situations in which,
due to particle-hole symmetry,
degenerate states can simultaneously resonate and therefore the more general
formula
(1) is required.

Consider now a N-NDOT-S structure, where the sub-space $B$ describes
the NDOT.
 At zero energy, if the isolated normal dot
 is on-resonance, then particle-hole symmetry ensures that
 a degeneracy occurs. Hence in this example,
the trace in equation (1) can be restricted to two terms and equation (2)
reduces to an expression involving 2x2 matrices. This yields
$$
G_{BB}\, =\,{1\over d}\left(\matrix{E-\epsilon_{-}
-\Sigma_{--}+\imath\Gamma_{--} &
\sigma^\prime_{+-}\cr
\sigma^\prime_{-+}&
E-\epsilon_+-\Sigma_{++}+\imath\Gamma_{++}\cr}\right)
\eqno{(3)},
$$
where $\epsilon_\pm$ are the particle and hole levels closest to the
quasi-particle energy $E$,
$
d\, =\,(E-\epsilon_+-\Sigma_{++}+\imath\Gamma_{++})
(E-\epsilon_{-}-\Sigma_{--}+
\imath\Gamma_{--})
-\vert\sigma^\prime_{+-}\vert^2
$
and we have written $\Sigma=\sigma'+\sigma$.
In the presence of a single normal lead, this yields for the electrical
conductance in units of $2e^2/h$ [14,27,28],
$$
G=2P^{-+}_{11}(E,H)
={8\Gamma_{++}\Gamma_{--}
\vert\sigma^\prime_{+-}\vert^2\over
\vert d\vert^2}
\eqno{(4)}
$$

First consider the zero energy limit (E=0),
where particle-hole symmetry implies that the two levels closest to
 $E=0$ satisfy
 $\epsilon_{-}\,=\,-\epsilon_+$ and therefore
a vanishing particle-level $\epsilon_+$ is accompanied by a degeneracy.
In this limit,
$\Sigma_{--}\,=\,-\Sigma_{++}$ and
$\Gamma_{--}\,=\,\Gamma_{++}$. Hence
$$
G
={8\Gamma^2_{++}
\sigma^\prime_{+-}\sigma^\prime_{-+}\over
((\epsilon_+ + \Sigma_{++})^2+\Gamma^2_{++} +
\vert\sigma^\prime_{+-}\vert^2)^2}
\eqno{(5)},$$
which was obtained in reference [10] for resonant transport at zero energy.
If $\vert\sigma^\prime_{+-}\vert^2=\Gamma^2_{++}$, then a resonance will
occur when $\epsilon_+ + \Sigma_{++}=0.$ Since this
involves only a single condition on the $\epsilon_+$, one expects resonances
to occur with approximately the same probability when the S-contact
is replaced by a N-contact.

At finite energies, this
result is drastically modified, because a resonance can now occur only
if both
$(E-\epsilon_+-\Sigma_{++}=0)$ and
$(E-\epsilon_{-}-\Sigma_{--}=0)$. The probability of simultaneously
satisfying both of these conditions is small and therefore we predict that
 the breaking of the particle-hole symmetry at $E \ne 0$ destroys
finite-voltage  conductance resonances.
{}From the form of the quartic energy denominator $\vert d\vert^2$,
the small number of surviving resonances will have a non-Lorentzian line-shape.
This is illustrated in figure 3, which shows
plots of the Andreev reflection coefficient $R_a=P^{-+}_{11}(E,H)$,
 for various values of coupling to the normal lead; figure 3(a)
has the strongest coupling and 3(d) the weakest.
All quantities are plotted as  functions of quasi-particle energy $E$.
Hence as well as a drastic reduction
in the probability of finding  finite-voltage resonances,
we predict that when a resonance does occur,
the usual Lorentzian line-shape
is replaced by a double peaked structure, with
relative peak heights determined by the difference between $\Gamma_{+}$
and $\Gamma_{-}$ at finite energies.
The above predictions are confirmed by
the results shown in figures 1 and 2, which were obtained from an exact
numerical solution of the Bogoliubov - de Gennes equation, as outlined in
reference 28. The inset of figure 2b shows numerical results for
the fine scale structure of a surviving resonance, in agreement with
the analytic prediction of figure 3.

When a composite N-S$S^\prime$ or N-NDOT-S$S^\prime$ structure is formed from
two or more superconductors, with different order parameter phases,
or from a single superconductor with an imposed phase gradient [36],
transport properties can be significantly modified if the
phase difference between two points is varied by $2\pi$[15-22].
In experimental realizations of such structures[23-26], the phase
difference between two superconducting contacts is modulated by
connecting the superconductors to a macroscopic, external superconducting
loop, whose phase is controlled by an applied magnetic field.
Sub-gap quasi-particles can penetrate only
a distance of order the superconducting
coherence length into the superconductor and therefore apart from
controlling the phase, the macroscopic loop plays no role in determining
the s-matrix of the region near the contacts.
 Since the electrical
conductance is a periodic function of the phase difference $\eta$, with period
$2\pi$ and since $\eta$ changes by $2\pi$ when the flux $\Phi$
through the macroscopic control-loop changes by a flux quantum $\Phi_0$,
such structures are galvanometric detectors of flux, with a sensitivity
comparable with that of a SQUID.
We now highlight generic properties of resonant interferometers
and predict that flux sensitivity can be significantly enhanced.

The starting point for this analysis is a sub-space $B$ containing
one or more superconductors, with eigenstates
$\vert f_\nu\rangle$ and eigenvalues $\epsilon_\nu$, which are periodic
functions of some dimensionless parameter $\eta$, with period $2\pi$.
In this example, since particle-hole degeneracy is lifted by intimate contact
with the superconductors, the trace in equation (1) reduces to
a single term and the electrical conductance takes the form
$$G(\eta)=
{8\Gamma_{\nu}(\eta+)\Gamma_{\nu}(\eta-)\over
\vert(E-\epsilon_{\nu}(\eta))-\Sigma_{\nu}(\eta)+\imath
\Gamma_{\nu}(\eta)\vert^2}
\eqno{(6)}.$$
Consider now the situation in which,
at $\eta=\eta_0$, the resonance condition
$E-\bar\Sigma_{\nu}(\eta_0)=0$ is satisfied,
where
$\bar\Sigma_{\nu}(\eta)=\epsilon_{\nu}(\eta)+\Sigma_{\nu}(\eta)$.
Then expanding equation (6) about $\eta_0$ yields
$$G(\eta)=
{8\Gamma_{\nu}(\eta_0+)\Gamma_{\nu}(\eta_0-)\over
[\partial\bar\Sigma_{\nu}(\eta_0)/\partial\eta_0]^2[\eta-\eta_0]^2+
\Gamma_{\nu}^2(\eta_0)}
\eqno{(7)}.$$
This demonstrates that with varying $\eta$, $G$ exhibits
a Lorentzian resonance of width
$\Gamma_\nu(\eta_0)/[\partial\bar\Sigma_{\nu}(\eta_0)/\partial\eta_0]$.

For the case $\eta=2\pi\Phi/\Phi_0$, noting that $\bar\Sigma_\nu(
\eta)$ can vary by at most an amount of order $\Delta_0$ as $\eta$
varies by $2\pi$,
yields an upper bound for
$[\partial\bar\Sigma_{\nu}(\eta_0)/\partial\eta_0]$ of order $\Delta_0/2\pi$.
Hence in terms of the flux through the external control loop,
the resonance width is greater than or of order
$\delta\Phi=2\pi\Phi_0\Gamma_\nu(\eta_0)/\Delta_0$.
For simplicity in the above analysis, we have considered only a single
 resonance and a normal lead with no closed channels; the latter merely
 shifts the position of the resonance, while the former may lead
 to the appearance of several resonances per flux quantum.
 If the temperature $T$ is greater than $\Gamma_\nu(\eta_0)/k_B$,
 then the resonance width will be of order
$\delta\Phi=2\pi\Phi_0k_BT/\Delta_0$.
For a device operating
at 1 Kelvin, formed from a cuprate  superconductor with a transition
temperature of 100 Kelvin, this yields $\delta\Phi\simeq\Phi_0/20$.

Finally we consider a superconducting dot with a uniform order
parameter, connected to two normal leads.
The eigenstates of a superconducting dot satisfy
$
{ H_B} \vert f_\nu\rangle\, =\,\epsilon_\nu\vert f_\nu\rangle
$
where ${ H_B}$ is the Bogoliubov-de Gennes operator for the isolated
dot.
For a dot with a
uniform real order parameter $\Delta_0$,
if
$\vert \phi\rangle$ is an eigenstate of the normal dot satisfying
$H_0\vert \phi\rangle\, =\,\epsilon^0_\nu\vert \phi\rangle$
then the solutions of the Bogoliubov equation  are of the form
$
\epsilon_\nu\, = \sqrt{(\epsilon^0_\nu)^2+\Delta_0^2}
$,
$
\vert f_\nu\rangle = \left(\matrix{\vert f^+_\nu\rangle\cr
\vert f^-_\nu\rangle}\right)
=\left(\matrix{u^+_\nu\vert \phi\rangle\cr
u^-_\nu\vert \phi\rangle}\right)
,$
where
$ \vert u^+_\nu\vert^2\, = (1 + {\epsilon^0_\nu/\epsilon_\nu})/2$ and
$\vert u^-_\nu\vert^2\, = (1 - {\epsilon^0_\nu/\epsilon_\nu})/2$.
At zero energy, this yields for the two-probe conductance
derived by Lambert[31,18],
$$
G\, =\,{4 \vert u^+_\nu\vert^2 \Gamma_1\Gamma_2\over\vert -\epsilon_\nu
+\sigma{\epsilon^0_\nu\over\epsilon_\nu} +
\imath (\Gamma_1+\Gamma_2)\vert^2}
\eqno{(8)},
$$
where $\Gamma_i$ is the broadening due to contact with lead $i$.

If $\delta\epsilon$ is the level spacing of the normal-state dot, then
equation (8) reveals
that  for
 $\vert\Delta_0\vert >\delta\epsilon\gg(\Gamma_1+\Gamma_2)$
 the contribution to the conductance from a single level is of order
$
G\, =\, {2\Gamma_1\Gamma_2\over \vert\Delta_0\vert^2}
$.
Hence for a large enough value of $\Delta_0/\delta\epsilon$, all resonances
will be suppressed
and switching on superconductivity will typically decrease $G$,
a behaviour which is well-known for N-S tunnel junctions.
However from the form of the denominator in equation (8), it is clear that
for small values of $\Delta_0$,
the switching-on of superconductivity can produce an
anomalous {\bf increase} in the conductance,
provided
$\sigma\, >\,\epsilon^0_\nu$. Since $\epsilon^0_\nu$ will
be randomly spread between $-{\delta\epsilon\over 2}$ and
${\delta\epsilon\over 2}$ this suggests
that the probability of a positive change is
approximately
${\sigma\over\delta\epsilon}$, a result which we have confirmed
through numerical solution of the Bogoliubov equation for such
structures[31].

We have presented a theoretical framework and general formulae
for resonant transport
through hybrid normal--superconducting nanostructures.
For  N-NDOT-S structures, we predict that finite-voltage resonances will be
almost
completely suppressed
by the switching on of superconductivity and those that survive
can have a double-peaked line-shape.
Such non-Lorentzian  resonances have been discussed in
other contexts [39] and  may generate non-exponential
delay-time curves. This destruction of resonances implies that
the ensemble averaged conductance  decreases with increasing
bias, a behaviour reminiscent of zero-bias anomalies  in the
sub-gap conductance of superconducting-semiconducting junctions[40,41].
We have also demonstrated that  N-SS$'$ structures   can possess
Lorentzian resonances on a scale much smaller than a flux quantum,
which suggests that these may provide a new class of magnetometers
with a sensitivity at least matching that of present-day SQUIDs.
Finally we have shown that switching on superconductivity in
 N-SDOT-N structures
can produce anomalous positive changes in the conductance.

\vskip 0.5truecm
\noindent
{\bf Acknowledgements.}
One of us (CJL) wishes to thank Markus B\"uttiker for fruitful
discussions and recent hospitality during the completion of
this manuscript.
This work is supported by the EPSRC, the EC HCM programme, the MOD,
the Institute for Scientific Interchange
and NATO.

\vfill
\eject

\vfill\eject
\noindent
$\underline {\hbox { Figure Captions.}}$
\vskip 0.5truecm
\noindent

\item {Figure 1.} For  quasiparticles of energy $E=0$,
the top graph shows the conductance $G$
when $\Delta_0\, =\,0$,  as a function of the mean diagonal element
$\epsilon_0$ of  a 2-dimension tight binding Hamiltonian
describing the NDOT. The band width of this Hamiltonian is 8
and the Fermi energy is $4-\epsilon_0$.
The lower graph shows corresponding results for
Andreev reflection coefficient $R_a$,
when the  order parameter
assumes a non-zero value.

\item {Figure 2.}
As for figure 1, except that the energy now takes a
non-zero, sub-gap value.
The inset of figure 2b shows the fine structure of a
typical surviving resonance.

\item {Figure 3.} Figures (a) to (d) show plots of equation (3) against
energy $E$, for decreasing values of the coupling to the normal lead.

\vfil\eject
\vskip 1.0truecm
\noindent
$\underline {\hbox {\bf References.}}$
\vskip 0.5truecm

\item{1.} Nanostuctures and mesoscopic systems, eds W.P.Kirk and M.A. Reed,
(Academic press, San Diego, 1992)

\item{2.} M. A. Kastner, Rev. Mod. Phys. {\bf 64} 849 (1992)

\item{3.} Single charge tunneling, eds. H. Grabert and M. Devoret
(Plenum Press, New york 1992)

\item{4.} V.N. Prigodin, K.B. Efetov and S. Iida, Phys. Rev. Lett {\bf 71}
1230 (1993)

\item{5.} H.U. Baranger and P.A. Mello, Phys. Rev. Lett. {\bf 73} 142 (1994)

\item{6.} R.A. Jalabert, J-L. Pichard and C.W.J. Beenakker, Europhys. Lett.
{\bf 27} 255 (1994)

\item{7.} T.M. Fromhold, M.L. Leadbeater, L. Eaves, T.J. Foster,
P.C. Main and F.W. Sheard, Surf. Sci. {\bf 305} 511 (1994)

\item{8.} T.M. Eiles, J.M. Martinis and M.H. Devoret, Phys. Rev. Lett.
{\bf 70} 1862 (1993)

\item{9} F. Hekking and Yu. Nazarov, Phys. Rev. Lett. {\bf } (1994)

\item{10.} C.W.J. Beenakker, cond-mat preprint (to appear in Mesoscopic Quantum
Physics, E. Akkermans, G.Montambaux and J.-L. Pichard, eds (North-Holland,
Amsterdam).)

\item{11.} V.A. Khlus, A.V.Dyomin and A.L. Zazunov, Physica {\bf C 214}
413-425 (1994).

\item{12.} G. Breit and E. P. Wigner, Phys. Rev. {\bf 49} 519 (1936).

\item{13.} L.D. Landau and E.M. Lifschitz, {\it Quantum mechanics
(Non-relativistic theory)} Pergamon Press, Oxford p. 603  (1977)

\item{14.} G.E. Blonder, M. Tinkham and T.M. Klapwijk, Phys. Rev. B. {\bf 25}
4515 (1982).

\item{15.} B.L. Al'tshuler and B.Z. Spivak, Sov. Phys. JETP
 {\bf 65}, 343 (1987).
\item{16.} H. Nakano and H. Takayanagi, Sol. St. Comm. {\bf 80} 997 (1991).
\item {17.} S. Takagi, Sol. St. Comm. {\bf 81} 579 (1992).
\item{18.}  C.J. Lambert  J.Phys. Condensed Matt. {\bf 5} 707 (1993).
\item {19.} V.C. Hui and C.J. Lambert, Europhys. Lett. {\bf 23}, 203 (1993).
\item {20.} F.W.J. Hekking and Yu.V. Nazarov, Phys. Rev. Lett. {\bf 71},
1625 (1993).
\item {21.} Yu.V. Nazarov, Phys. Rev. Lett. {\bf 73},
1420 (1994).
\item {22.} A.V. Zaitsev, Phys.Lett. {\bf A194}, 315 (1994).

\item{23.} B.J. van Wees, A. Dimoulas, J.P. Heida, T.M. Klapwjk,
W.v.d. Graaf, and G. Borghs, Physica {\bf B203}, 285 (1994).
\item{24.} P.G.N. de Vegvar, T.A. Fulton, W.H. Mallison,  and R.E. Miller,
 Phys. Rev. Lett. {\bf 73}, 1416 (1994).
\item{25.} H. Pothier, S. Gueron, D. Esteve, and M.H. Devoret,
Physica {\bf B203}, 226 (1994).
\item{26.} V.T. Petrashov, V.N. Antonov, S.V. Maksimov and
R. Sh. Shaikhaidarov, Sov. Phys. JETP Lett. {\bf 59} 551 (1994)

\item{27.} C.J. Lambert, J. Phys.: Condens. Matter, {\bf 3} 6579 (1991).

\item{28.} C.J. Lambert, V.C. Hui and S.J. Robinson, J.Phys.: Condens. Matter,
{\bf 5} 4187 (1993).

\item{29.} C.J. Lambert, Physica {\bf B 203} 201 (1994)

\item{30.} Y. Takane and H. Ebisawa, J. Phys. Soc. Jap. {\bf 61} 1685 (1992).

\item{31.} N.R. Claughton, M. Leadbeater and C.J. Lambert,
to be published

\item{32.} M. B\"uttiker, Y. Imry and M.ya Azbel, Phys. Rev. {\bf A 30} 1982
(1982)

\item{33.} M. B\"uttiker, Phys. Rev. {\bf B 38} 12724 (1988)

\item{34.} J.U. N\"ockel and A.D. Stone, Phys. Rev. {\bf B 50} 17415 (1995)

\item{35.}  M. B\"uttiker, IBM J. Res. Dev. {\bf 32} 63 (1988)

\item{36.} P.M.A. Cook, V.C. Hui and C.J. Lambert, Euro. Phys. Lett. {\bf 30}
355 (1995)

\item{37.} V.C. Hui and C.J. Lambert, J. Phys. Condens. Matter {\bf 5}
L651 (1993).

\item{38.} V.C. Hui and C.J. Lambert, Physica {B194-196} 1673 (1994).

\item{39.} M. L. Goldberger and K. M. Watson, Phys. Rev. {\bf 136} B1472
(1964).

\item{40.} A. Kastalasky, A.W. Kleinsasser, L.H. Greene, R. Bhat and
J.P. Harbison, Phys. Rev. Lett. {\bf 67} 3026 (1991)

\item{41.} B.J. van Wees, P. de Vries, P. Magnee and T.M. Klapwijk,
Phys. Rev. Lett. {\bf 69} 510 (1992)

\end